\documentclass[twocolumn,aps,prl,amsmath,amssymb,showpacs,preprintnumbers]{revtex4}
\usepackage{graphicx}
\begin{document}


\title{Testing non-associative quantum mechanics}

\author{Martin Bojowald}
\email{bojowald@gravity.psu.edu}
\affiliation{Institute for Gravitation and the Cosmos,
The Pennsylvania State University, 104 Davey Lab, University Park,
PA 16802, USA}

\author{Suddhasattwa Brahma}
\email{sxb1012@psu.edu}
\affiliation{Institute for Gravitation and the Cosmos,
The Pennsylvania State University, 104 Davey Lab, University Park,
PA 16802, USA}

\author{Umut B\"{u}y\"{u}k\c{c}am}
\email{uxb101@psu.edu}
\affiliation{Institute for Gravitation and the Cosmos,
The Pennsylvania State University, 104 Davey Lab, University Park,
PA 16802, USA}

\pacs{03.65.Sq, 03.65.Fd, 14.80.Hv, 02.10.De}

\begin{abstract}
  The familiar concepts of state vectors and operators in quantum mechanics
  rely on associative products of observables. However, these notions do not
  apply to some exotic systems such as magnetic monopoles, which have long
  been known to lead to non-associative algebras.
Their quantum
  physics has remained obscure. This letter presents the first derivation of
potentially
  testable physical results in non-associative quantum mechanics, based on
  effective potentials. They imply new effects which cannot be mimicked in
  usual quantum mechanics with standard magnetic fields.
\end{abstract}

\maketitle

Quantum mechanics is being tested ever more precisely by experiments, even
while conceptual questions remain. We suggest a new kind of test in an
extended setting in which the usual concepts of wave functions or state
vectors and operators do not exist. Therefore, the standard axioms about
outcomes of individual measurements are unavailable, or at least not known
yet, and even at a practical level, no computational methods have been
available so far. Here we show how one can derive semiclassical corrections to
the motion of a particle and associated new phenomena.

The formalism of state vectors and operators implies that the action of the
latter on the former is associative:
\begin{equation}
(\hat{A}\hat{B})\hat{C}\psi=
\hat{A}\hat{B}\psi'= \hat{A}(\hat{B}\psi')=\hat{A}(\hat{B}\hat{C})\psi
\end{equation}
if $\psi'=\hat{C}\psi$, for an arbitrary $\psi$.  However, some exotic
ingredients, such as magnetic monopoles, require an underlying non-associative
algebra in order to quantize such systems. Quantum observables then no longer
obey $(\hat{A}\hat{B})\hat{C}=\hat{A}(\hat{B}\hat{C})$. Not much has been
known about physical effects when this basic identity is not available. In
this letter, we develop and utilize a novel method in order to reveal testable
quantum effects in such a system.

A non-associative algebra cannot be represented by operators on a Hilbert
space. Instead, one has to work with an abstract non-associative algebra,
which can be constructed by methods of deformation quantization as applied in
\cite{NonGeoNonAss,MSS1,BakasLuest,MSS2,MSS3}. States in this setting are not
defined as normalized vectors in a Hilbert space, but as suitable linear
functionals $\hat{A}\mapsto\langle\hat{A}\rangle$ from the algebra of
observables to the complex numbers. (For associative observable algebras, this
definition of a state is equivalent to the Hilbert-space picture thanks to
the Gelfand--Naimark--Segal theorem; see for instance \cite{LocalQuant}.)  The
primary object is therefore not a wave function but the set of expectation
values assigned by a single state to all possible observables. We will
demonstrate, for the first time in the context of non-associative quantum
mechanics, that algebraic properties of such expectation-value functionals can
be used to derive new semiclassical effects.

The best-known example of a non-associative quantum system is a charged
particle in a magnetic monopole density \cite{MagneticCharge}. Even if
fundamental magnetic monopoles may not exist, such systems are gaining
interest from a physical perspective, following recent constructions of analog
systems of magnetic monopoles in condensed-matter physics
\cite{MonopoleIce1,MonopoleIce,MonopoleCharge,DiracMon,MonopoleCurrent,MonopoleLattice}. (Related
models also play a role in string theory
\cite{DualDoubled,NonAssGrav,NonGeoNonAss,TwistedNonAss,NonAssDef}.) As is
well known, the canonical momentum of a charged particle in a magnetic field
$\vec{B}$ without monopoles, ${\rm div}\vec{B}=0$, is a combination of the
particle velocity and the vector potential.  For the kinematical momentum
$\vec{p}=m\dot{\vec{q}}$, one has non-canonical commutators of momentum
components,
\begin{equation} \label{pp}
 \left [\hat{p}_j,\hat{p}_k\right]=i e\hbar   \sum_{l=1}^3\epsilon_{jkl}
 \hat{B}^l
\end{equation}
where $e$ is the particle's electric charge.

This relation depends only on the magnetic field and does not require a vector
potential. Therefore, it can be used to define the basic commutators of a
charged particle moving in a magnetic field with ${\rm div}\vec{B}\not=0$.
Regarding quantization, such fields can be of two types. If ${\rm
  div}\vec{B}\not=0$ at isolated points of Dirac monopoles, these points can
be excised and standard quantum mechanics applies. We are interested in the
second type of charges not subject to Dirac quantization, allowing us to
include fields with continuous magnetic charge densities for which no vector
potential exists.
The resulting algebra then does not fulfill the Jacobi identity of
commutators \cite{JackiwMon,Malcev}:
\begin{eqnarray}
&& [[\hat{p}_x,\hat{p}_y],\hat{p}_z]+  [[\hat{p}_y,\hat{p}_z],\hat{p}_x]+
 [[\hat{p}_z,\hat{p}_x],\hat{p}_y]\nonumber \\
&=& ie\hbar  \sum_{j=1}^3[\hat{B}^j,\hat{p}_j]= -e\hbar^2\widehat{{\rm
    div}\vec{B}}\,.
\end{eqnarray}
As the name suggests, the Jacobi {\em identity} normally follows without
further assumptions, provided the algebra is associative. The non-zero result
just obtained can therefore be consistent only if the multiplication of
momentum components is not associative. (Finite translations generated by
momentum operators are not associative \cite{Jackiw}. The classical analog is
a twisted Poisson bracket \cite{TopoTwist,WZWTwist,Twisted}.)


A physical version of the property of non-associativity is a ``triple''
uncertainty relation, just as the usual uncertainty relation is a consequence
of non-commuting operators. As usual, (\ref{pp}) implies that $\Delta p_x
\Delta p_y\geq \frac{1}{2} e\hbar \langle\hat{B}^z\rangle$: a large magnetic
field in the $z$-direction deflects the particle from a straight line, making
it harder to measure momentum components. A characteristic of monopole fields
is that they change along the direction in which they point, for instance if
$B^z=\mu z$ with a constant $\mu$. The commutator of $\hat{p}_x$ and
$\hat{p}_y$ then depends on the measurement of $\hat{z}$, which itself is
subject to the standard uncertainty relation $\Delta z\Delta p_z\geq
\frac{1}{2}\hbar$. Therefore, all three fluctuations, $\Delta p_x$, $\Delta
p_y$ and $\Delta p_z$, together determine how small the momentum fluctuations
can be.

As already mentioned, states then cannot be defined as vectors in a Hilbert
space, but their physical properties can be analyzed by treating them as
linear expectation-value functionals on the algebra. Any such functional must
be normalized, $\langle\hat{\mathbb I}\rangle=1$ for the identity
$\hat{\mathbb I}$ in the algebra, and obey a positivity condition which
implies uncertainty relations. Having identified basic operators as the
components of position and kinematical momentum, we can parameterize a state
by the basic expectation values $\langle\hat{q}_i\rangle$ and
$\langle\hat{p}_j\rangle$ as well as fluctuations, correlations and higher
moments. The latter are defined to be of the form
\begin{equation} \label{DeltaDef}
 \Delta(p_ip_j):= \frac{1}{2}
 \langle\hat{p}_i\hat{p}_j+\hat{p}_j\hat{p}_i\rangle- \langle\hat{p}_i\rangle
 \langle\hat{p}_j\rangle
\end{equation}
for the example of two momentum components. (With this notation, we slightly
modify the usual denotation of quantum fluctuations, identifying
$\Delta(p_i^2)=(\Delta p_i)^2$.) The symmetrization in (\ref{DeltaDef}) takes
into account the non-commuting nature of kinematical momentum components in a
magnetic field. For higher moments with more than two factors of momentum or
position components, different symmetrizations are possible, of which we
choose, following \cite{EffAc}, totally symmetric (or Weyl) ordering by
summing with equal weights over all permutations of factors. Moreover,
non-associativity requires a fixed choice for the bracketing of products of
observables, which we choose to be done from the left as in \cite{NonAss}.

A Hamiltonian leads to equations of motion for the basic expectation values
coupled to moments, giving an infinite-dimensional dynamical system. In a
semiclassical approximation, only a finite number of moments need be
considered, corresponding to a specific order in $\hbar$.  The Hamiltonian we
use in this letter is of the standard form,
\begin{equation}
 \hat{H}=\frac{1}{2m} \sum_{j=1}^3\hat{p}_j^2+V(\hat{x},\hat{y},\hat{z})\,,
\end{equation}
where interactions of a charged particle with the magnetic field are
implemented not by a term in the potential but by the non-trivial
commutators of momentum components. (The potential $V(x,y,z)$ for an
additional, non-magnetic force will be specified below.) Given a Hamiltonian,
equations of motion for expectation values and moments follow by using
\begin{equation}\label{EOMgenerate}
 \frac{{\rm d}\langle\hat{O}\rangle}{{\rm d}t} =
 \frac{\langle[\hat{O},\hat{H}]\rangle}{i\hbar}
\end{equation}
which is still available in the non-associative case. However, the
non-associative nature requires great care when evaluating commutators of
products of the basic observables, for which we refer to \cite{NonAss}.

The specific effect we will derive, related to stable motion in an effective
potential, requires the particle to be completely confined.  A magnetic field
in the $z$-direction, $B^x=0=B^y$, confines the particle motion to a plane
normal to the magnetic field. For complete confinement, we combine the
magnetic force with a harmonic force in the same direction, choosing the
potential to be $V(x,y,z)=\frac{1}{2}m\omega^2z^2$. This force could be
generated by an electric field. For simplicity, we consider a linear
$z$-component $B^z=\mu z$, so that $\mu$ is the magnetic charge density. The
resulting Hamiltonian is
\begin{equation}
 \hat{H}=\frac{1}{2m} \sum_{j=1}^3\hat{p}_j^2+\frac{1}{2}m\omega^2\hat{z}^2\,,
\end{equation}
and the magnetic field enters via $[\hat{p}_x,\hat{p}_y]=ie\mu\hbar \hat{z}$
while the other pairs of momentum components commute.

We are interested in deriving an effective potential for the motion of such a
particle. If one knows a suitable state of the system, the effective potential
can be obtained from the expectation value of the Hamiltonian in which one
sets all momentum expectation values to zero in order to remove the kinetic
term, $V_{\rm eff}=\langle\hat{H}\rangle_{\langle\hat{p}_i\rangle=0}$. (We do
not require solutions for momentum expectation values to be zero at all
times.)  In the given case with a quadratic Hamiltonian, the effective
potential is the classical potential plus a sum of fluctuations:
\begin{eqnarray} \label{Veff}
V_{\rm eff}(\langle\hat{z}\rangle) &=&
\frac{1}{2}m\omega^2\langle\hat{z}\rangle^2\\
&&+ \frac{1}{2m}
\left(\Delta(p_x^2)+ \Delta(p_y^2)+\Delta(p_z^2)\right) +
\frac{1}{2}m\omega^2\Delta(z^2)\,. \nonumber
\end{eqnarray}
In order to express this potential as a function of the coordinates, we have
to compute the values of quantum fluctuations. Following the methods of
\cite{CW}, we can compute the relevant state properties without using a wave
function. Instead, we solve equations of motion for fluctuations in an
adiabatic approximation (giving stationary moments in a near-coherent state)
and saturating uncertainty relations (minimizing fluctuations). For
well-understood (associative) systems such as anharmonic oscillators
\cite{EffAcQM} or Coleman--Weinberg potentials in self-interacting scalar
field theories \cite{ColemanWeinberg}, the correct results are obtained in
this way \cite{EffAc,CW}. In our new situation, we minimize fluctuations by
saturating the uncertainty relations, in the standard form for $q_i$ and $p_i$
and for non-commuting momentum components.

We can derive Ehrenfest-type equations of motion by using (\ref{EOMgenerate})
for the moments. (See also \cite{NonAss}.)  Expanded up to first order in
$\hbar$ for semiclassical states, thus including no moments of higher than
second order, they are
\begin{eqnarray} \label{qiqj}
m\dot{\Delta}(q_iq_j)&=&\Delta(p_xq_i)\delta_{jx} +\Delta(p_xq_j)\delta_{ix}
+\Delta(p_yq_i)\delta_{jy} \nonumber\\
&&  +\Delta(p_yq_j)\delta_{iy}+\Delta(p_zq_i)\delta_{jz} +\Delta(p_zq_j)\delta_{iz}
\end{eqnarray}
for all position moments,
\begin{eqnarray}
m\dot{\Delta}(p_xq_i) &=& \Delta(p_x^2)\delta_{ix}+ \Delta(p_xp_y)\delta_{iy}+
\Delta(p_xp_z)\delta_{iz} \nonumber\\
&&+ e\mu \left(\langle\hat{z}\rangle\Delta(p_yq_i)-
\langle\hat{q}_i\rangle \Delta(p_yz)\right)\,,\label{pxqi}\\
m\dot{\Delta}(p_yq_i) &=& \Delta(p_xp_y)\delta_{ix}+ \Delta(p_y^2)\delta_{iy}+
\Delta(p_yp_z)\delta_{iz}\nonumber\\
&&- e\mu \left(\langle\hat{z}\rangle\Delta(p_xq_i)-
\langle\hat{q}_i\rangle \Delta(p_xz)\right)\,, \label{pyqi}\\
m\dot{\Delta}(p_zq_i) &=& \Delta(p_xp_z)\delta_{ix}+ \Delta(p_yp_z)\delta_{iy}+
\Delta(p_z^2)\delta_{iz}\nonumber\\
&& -m^2\omega^2 \Delta(q_iz)\label{pzqi}
\end{eqnarray}
for the position-momentum covariances,
\begin{eqnarray}
m \dot{\Delta}(p_xp_y) &=& -e\mu \left(\langle\hat{z}\rangle\Delta(p_x^2)-
  \langle\hat{z}\rangle\Delta(p_y^2)\right.\nonumber\\
&&\left.-\langle\hat{p}_x\rangle\Delta(p_xz)+
\langle\hat{p}_y\rangle\Delta(p_yz)\right)\,, \label{pxpy}\\
m \dot{\Delta}(p_yp_z) &=& -e\mu \left(\langle\hat{z}\rangle\Delta(p_xp_z)+
\langle\hat{p}_x\rangle\Delta(p_zz)\right)\nonumber\\
&&-m^2\omega^2\Delta(p_yz)\,,\label{pypz}\\
m \dot{\Delta}(p_xp_z) &=& e\mu \left(\langle\hat{z}\rangle\Delta(p_yp_z)+
\langle\hat{p}_y\rangle\Delta(p_zz)\right)\nonumber\\
&&-m^2\omega^2\Delta(p_xz) \label{pxpz}
\end{eqnarray}
for momentum covariances, and
\begin{eqnarray}
m\dot{\Delta}(p_x^2) &=& 2e\mu
\left(\langle\hat{z}\rangle\Delta(p_xp_y)+
2\langle\hat{p}_y\rangle\Delta(p_xz)+
\langle\hat{p}_x\rangle\Delta(p_yz)\right)\, \label{px2}\\
m\dot{\Delta}(p_y^2) &=& -2e\mu
\left(\langle\hat{z}\rangle\Delta(p_xp_y)+
\langle\hat{p}_y\rangle\Delta(p_xz)+
2\langle\hat{p}_x\rangle\Delta(p_yz)\right)\, \label{py2}\\
m\dot{\Delta}(p_z^2) &=& -2m^2\omega^2 \Delta(zp_z) \label{pz2}
\end{eqnarray}
for momentum fluctuations.

We solve these equations to zeroth adiabatic order in the moments, so that all
time derivatives on the left-hand sides can be set to zero. The moments are
then subject to linear algebraic equations.  In order to solve the set of
coupled equations, we use (\ref{qiqj}) for all possible index combinations to
conclude that
\begin{eqnarray}
 \Delta(xp_x)=\Delta(yp_y)=\Delta(zp_z)&=&0 \label{A1}\\
 \Delta(yp_x)+\Delta(xp_y)&=&0 \label{A2}\\
\Delta(zp_x)+\Delta(xp_z)&=&0 \label{A3}\\
\Delta(zp_y)+\Delta(yp_z)&=&0 \label{A4}\,.
\end{eqnarray}
Using equations for mixed position-momentum moments, we obtain
\begin{eqnarray}
 \Delta(p_x^2) &=& -e\mu\left(\langle\hat{z}\rangle\Delta(xp_y)-
\langle\hat{x}\rangle\Delta(zp_y)\right)\,, \label{B1}\\
  \Delta(p_xp_y) &=&
  -e\mu\left(\langle\hat{z}\rangle\Delta(yp_y)-
\langle\hat{y}\rangle\Delta(zp_y)\right)\,,\label{B2}\\
 \Delta(p_xp_z) &=& 0\label{B3}
\end{eqnarray}
from (\ref{pxqi}),
\begin{eqnarray}
 \Delta(p_xp_y) &=& e\mu \left(\langle\hat{z}\rangle\Delta(xp_x)-
\langle\hat{x}\rangle\Delta(zp_x)\right)\,,\label{C1}\\
 \Delta(p_y^2) &=& e\mu \left(\langle\hat{z}\rangle\Delta(yp_x)-
\langle\hat{y}\rangle\Delta(zp_x)\right)\,,\label{C2}\\
 \Delta(p_yp_z) &=& 0 \label{C3}
\end{eqnarray}
from (\ref{pyqi}), and
\begin{eqnarray}
 \Delta(p_xp_z) &=& m^2\omega^2\Delta(xz)\,, \label{D1}\\
 \Delta(p_yp_z) &=& m^2\omega^2\Delta(yz)\,, \label{D2}\\
 \Delta(p_z^2) &=& m^2\omega^2\Delta(z^2) \label{D3}
\end{eqnarray}
from (\ref{pzqi}).  The equations of motion (\ref{pxpy}), (\ref{pypz}) and
(\ref{pxpz}) for momentum covariances provide
\begin{eqnarray}
 \Delta(p_x^2)-\Delta(p_y^2) &=&
 \frac{\langle\hat{p}_x\rangle\Delta(zp_x)-
\langle\hat{p}_y\rangle\Delta(zp_y)}{\langle\hat{z}\rangle}\,, \label{E1}\\
 \Delta(zp_z) &=& -\frac{m^2\omega^2}{e\mu \langle\hat{p}_x\rangle} \Delta(zp_y)-
 \frac{\langle\hat{z}\rangle}{\langle\hat{p}_x\rangle}
\Delta(p_xp_z)\,, \label{E2}\\
 \Delta(zp_z) &=& \frac{m^2\omega^2}{e\mu \langle\hat{p}_y\rangle}
\Delta(zp_x)- \frac{\langle\hat{z}\rangle}{\langle\hat{p}_y\rangle}
 \Delta(p_yp_z)\,. \label{E3}
\end{eqnarray}

Since $\Delta(zp_z)=0$ from (\ref{A1}) and $\Delta(p_xp_z)=0=\Delta(p_yp_z)$
from (\ref{B3}) and (\ref{C3}), (\ref{E2}) and (\ref{E3}) imply
$\Delta(zp_x)=0=\Delta(zp_y)$.  From (\ref{E1}) and (\ref{B2}) or (\ref{C1}),
we immediately conclude that
\begin{equation} \label{pxpyrel}
 \Delta(p_x^2)=\Delta(p_y^2) \quad\mbox{and}\quad \Delta(p_xp_y)=0 \,,
\end{equation}
also using (\ref{A1}). These values are consistent with (\ref{B1}) and
(\ref{C2}), in which the same fluctuations appear. All equations are then
solved and the adiabatic approximation is self-consistent, showing
that an effective potential exists.

We now consider states saturating the uncertainty relations. For the pair
$(z,p_z)$, we have the standard one,
\begin{equation}
 \Delta(z^2)\Delta(p_z^2)-\Delta(zp_z)^2\geq \frac{\hbar^2}{4}\,,
\end{equation}
while (\ref{pp}) implies an uncertainty relation
\begin{equation}
  \Delta(p_x^2)\Delta(p_y^2)-\Delta(p_xp_y)^2\geq \frac{1}{4}
  e^2\hbar^2\langle\hat{B}\rangle^2\,.
\end{equation}
If both inequalities are saturated, we obtain
\begin{equation}
 \Delta(p_x^2)=\Delta(p_y^2)=\frac{1}{2}e\hbar \langle\hat{B}\rangle
\end{equation}
and
\begin{equation}
 \Delta(z^2)= \frac{\hbar}{2m\omega}\quad,\quad \Delta(p_z^2) =
 \frac{1}{2}\hbar\omega\,.
\end{equation}
Finally, inserting these values in (\ref{Veff}), we obtain
\begin{equation} \label{VeffSol}
 V_{\rm eff}(\langle\hat{z}\rangle) =
 \frac{1}{2}m\omega^2\langle\hat{z}\rangle^2 + \frac{1}{2}\hbar
 \frac{eB(\langle\hat{z}\rangle)}{m}+\frac{1}{2}\hbar\omega\,.
\end{equation}

If the magnetic field is constant, the fraction $eB/m=\omega_{\rm c}$ in
(\ref{VeffSol}) is the cyclotron frequency. It is well known that the
Hamiltonian of a charged particle in a constant magnetic field can be
transformed to one of a harmonic oscillator with the cyclotron frequency, so
that our derivation provides the correct result of a constant $\hbar$-term in
the effective potential given by the sum of zero-point energies
$\frac{1}{2}\hbar\omega_{\rm c}$ and $\frac{1}{2}\hbar \omega$ of two
uncoupled oscillators.

In the case of a magnetic field with constant charge density, the effective
potential is linear, implying a new constant force from quantum effects. The
force
\begin{equation}
 F_{\rm eff}= -\frac{e\mu\hbar}{2m}
\end{equation}
points in the direction of the magnetic field, or in the same direction in
which the harmonic force is acting.  We can combine the classical potential
$\frac{1}{2}m\omega^2z^2$ with the linear quantum potential $\frac{1}{2}
e\mu\hbar z/m$ and write the effective potential as
\begin{equation}
 V_{\rm eff}(\langle\hat{z}\rangle)=\frac{1}{2}m\omega^2
 \left(\langle\hat{z}\rangle+\frac{1}{2}\frac{e\mu\hbar}{m^2\omega^2}\right)^2
 +\frac{1}{2}\hbar\omega\,.
\end{equation}
(Rewriting (\ref{VeffSol}) in this way generates an $\hbar^2$-term, which we
do not include here because all our derivations were up to first order in
$\hbar$. The $\hbar^2$-term can be absorbed in the next order corrections
coming from higher moments.) The minimum of the harmonic potential is shifted
by
\begin{equation}
  \delta z = -\frac{1}{2}\frac{e\mu\hbar}{m^2\omega^2}
\end{equation}
as the main first-order quantum effect. Interestingly, the shift is inversely
proportional to $\omega$, so that a small frequency can enlarge the quantum
effect.  (For $\omega\to0$, the shift diverges. However, it is then simply
ill-defined because the harmonic potential disappears in the limit and does
not distinguish a center in the $z$-direction.) Also, as may be expected for
an effect of quantum back-reaction, the shift is larger for particles with
smaller mass.

In addition to harmonic oscillations around a shifted center, the charged
particle would move along a circle in the $x-y$-plane as a consequence of the
non-zero $B^z$ for $z\not=0$. Classically, without the shift in the harmonic
potential, stable circular motion would not be possible because at $z=0$,
where the harmonic force vanishes, the magnetic field is zero. Given the shift
of the minimum, stable circular motion is now possible with the particle
confined to move in a plane at fixed $z=\delta z$.

This effect cannot be mimicked by magnetic fields without monopole densities.
Such a magnetic field could produce a $z$-dependent potential only if there
are non-vanishing components $B^x$ or $B^y$, either by cancelling the
$z$-derivative of $B^z$ in the divergence or by having the $z$-dependence come
only from $B^x$ or $B^y$.  However, the motion would then be more complicated
than circular motion in the $x-y$-plane at some fixed value of $z$.

Our results have important conceptual and potentially observable
consequences. They demonstrate that physical effects can be derived in quantum
mechanics even when the usual and widely used notions of state vectors and
operators are unavailable. Non-associative quantum mechanics is thereby shown
to be meaningful physically, which, despite its exotic appearance, can be
applied in diverse ways, including some versions of string theory and analog
magnetic monopoles.

Regarding the latter, we have specialized our general methods to a system in
which closed-form solutions can be obtained, providing a model system with
clear new effects. Such models always play important roles in situations like
the present one: not much is known about testable quantum effects of analog
condensed matter monopoles, even while experimental realizations seem to be
within reach \cite{QuantMonopole}.  Our model amounts to an idealized example
which brings out new effects clearly. 

In practice, although it seems hard to have a constant monopole density, for
sufficiently large amplitude of the oscillating motions of a charged particle,
it is conceivable that
a fine lattice of magnetic monopoles could be used to test the new effect
found here.
Specifically, one should arrange the lattice in cylinder shape, so as to
impose a preferred direction identified here with the $z$-direction. On scales
larger than the lattice spacing (but well within the entire lattice), the
complicated dynamics of electric charges moving around monopoles can be
approximated by electric charges moving through a uniform monopole density 
to which our methods apply.
Analog monopoles do have Dirac strings \cite{QuantMonopole}, which may still
have an effect after averaging to a continuous density, making the magnetic
field non-linear.  For more accurate derivations of the effective potential,
applying our methods to non-linear magnetic fields, the same equations for
moments are available, but they are coupled in more complicated ways which are
likely to require numerical input and further research.
Similarly, the equations can be extended to higher orders in $\hbar$ by
including higher moments, but again we are not aware of closed analytic
solutions.

\smallskip

\noindent {\bf Acknowledgements:}
This work was supported in part by NSF grant PHY-1307408.


\bigskip

\noindent {\bf Erratum: Testing non-associative quantum mechanics
 [Phys.\ Rev.\ Lett.\
  115, 220402 (2015)]}

\medskip

This Letter corrects a sign mistake made in solving an uncertainty relation
for minimal momentum fluctuations, and discusses implications for the
resulting physical effect. In \cite{NonAssEffPot}, we first solved
semiclassical equations of motion for
stationary second-order moments, and then imposed minimal uncertainty by
saturating uncertainty relations. This led us to the equations
$\Delta(p_x^2)=\Delta(p_y^2)$, $\Delta(p_xp_y)=0$ and
\begin{equation}
   \Delta(p_x^2)\Delta(p_y^2)-\Delta(p_xp_y)^2\geq \frac{1}{4}
  e^2\hbar^2\langle\hat{B}\rangle^2
\end{equation}
which we saturated by $\Delta(p_x^2)=\frac{1}{2} e\hbar
\langle\hat{B}\rangle$. However, the magnetic field can be negative, and
therefore we should have written $\Delta(p_x^2)=\frac{1}{2} e\hbar
|\langle\hat{B}\rangle|$ (assuming $e>0$). These fluctuations enter the
effective potential which now reads
\begin{equation} \label{Veff}
 V_{\rm eff}(\langle\hat{z}\rangle) =
 \frac{1}{2}m\omega^2\langle\hat{z}\rangle^2 + \frac{1}{2}\hbar
 \frac{e|B(\langle\hat{z}\rangle)|}{m}+\frac{1}{2}\hbar\omega\,.
\end{equation}

In our specific example, $B(\langle\hat{z}\rangle)=\mu \langle\hat{z}\rangle$
appeared to lead to a shift of the minimum of the effective potential by
quantum effects. However, with the absolute value in (\ref{Veff}), the
effective potential is still reflection symmetric around
$\langle\hat{z}\rangle=0$, and there is no shift of the minimum. Instead, the
effective potential differs from the classical potential by a kink around
$\langle\hat{z}\rangle=0$, accompanied by a modification of the potential in a
neighborhood around $\langle\hat{z}\rangle=0$ where the linear contribution to
the potential is dominant. (At the kink, where the $\hbar$-correction implied
by the magnetic field is zero, it is likely that higher-order corrections are
relevant.) The motion of a charged particle in a magnetic monopole density
therefore differs from the classical motion by anharmonic behavior around the
minimum of an additional quadratic potential.

In summary, our main result of giving a physical interpretation to
non-associative quantum mechanics and providing new methods to make
effects calculable remains unchanged. However, in terms of potential
observations in the specific example provided by us, this implies that
one would have to detect deviations from harmonic motion rather than a
shift of the equilibrium position, as reported earlier.

\noindent {\bf Acknowledgements:}
This work was supported in part by NSF grant PHY-1307408.


\end{document}